\documentclass[twocolumn, showpacs,superscriptaddress]{revtex4}
\usepackage{epsfig}
\usepackage{amssymb}
\usepackage{amsmath}
\usepackage{amsbsy}
\usepackage{color}
\usepackage{ulem}
\usepackage{epstopdf}

\begin{document}
\title{Ring polymers in melts and solutions: scaling and crossover}
\author{Takahiro Sakaue}
\email{sakaue@phys.kyushu-u.ac.jp}
\affiliation{Department of Physics, Kyushu University 33, Fukuoka 812-8581, Japan}
\affiliation{PRESTO, Japan Science and Technology Agency (JST), 4-1-8 Honcho Kawaguchi, Saitama 332-0012, Japan}

\begin{abstract}
We propose a simple mean-field theory for the structure of ring polymer melts. By combining the notion of topological volume fraction and a classical van der Waals theory of fluids, we take into account many body effects of topological origin in dense systems. We predict that although the compact statistics with the Flory exponent $\nu=1/3$ is realized for very long chains, most practical cases fall into the crossover regime with the apparent exponent $\nu = 2/5$ during which the system evolves toward a topological dense-packed limit.  
\end{abstract}


\maketitle

Statistics of melts and concentrated solutions of ring polymers is a longstanding problem in polymer physics~\cite{C-D,Khokhlov_Nechaev,crumpled_globule, Ring_linear_compatibility,Obukhov,Arrighi,Takano,Pakula,Muller1, Brown, Muller2, Muller3,Vettorel,Suzuki}.
Unlike linear chain systems, their physical properties crucially depends on the preparation history during which the topology of the system is frozen.
Then, the non-crossing requirement creates {\it topological constraints} which impose nontrivial restrictions on the phase space of the system, hence, have a strong influence even in statics.

%For example, if the chain ends are closed in a narrow space, a resultant ring may contain rather complex self-knotting as is the case of phage DNA in virus capsid. On the other hand, if the end-closure reaction takes place in the melt of long linear polymers, it would end up with massively linked rings with a high degree of concatenation.

In the linear polymer counterpart, a well-known Flory theorem states that the chain conformation is Gaussian characterized by the Flory exponent $\nu=1/2$ due to the screening of excluded-volume interactions~\cite{deGennes, Doi_Edwards, Grosberg_Khokhlov}. This simple (but surprising) result, however, no longer holds for the ring polymer melt.
The most basic question arises in the melt of ring polymers free from any mutual-linking and self-knotting.
There have been several experimental~\cite{Arrighi, Takano} and numerical studies~\cite{Pakula,Muller1, Brown, Muller2, Muller3, Vettorel,Suzuki} in this direction, but the clear-cut conclusion has not been attained yet. 

In their seminal paper, Cates and Deutsch (C-D) argued that the unconcatenated rings in the melt may have statistics intermediate between those of collapsed ($\nu=1/3$) and Gaussian ($\nu=1/2$) chains~\cite{C-D}. Specifically, they proposed a conjecture on the scaling exponent $\nu=2/5$ based on Flory-type mean-field theory.
While this leading theoretical guide seems to be supported by following numerical simulations~\cite{Pakula,Muller1, Brown, Muller2, Muller3}, some of more recent observations claim the collapsed statistics $\nu=1/3$ for sufficiently long rings~\cite{Vettorel,Suzuki}.
The latter result was hypothesized by Khokhlov and Nechaev based on the analogy with lattice animals~\cite{Khokhlov_Nechaev}. Closely related to this is the crumpled globule (CG) model~\cite{crumpled_globule, Ring_linear_compatibility} which was originally proposed by Grosberg, et. al. as a long-lived kinetic intermediate on the pathway of the linear chain collapse~\cite{crumpled_globule}, and then, hypothesized as a large scale DNA organization in interphase chromosomes. 
This suggests an interesting link between polymer topology and biologically important problems such as the formation of the chromosome territories~\cite{chromosome_territories}.

It is important to keep in mind that the topological effect manifests itself in the scale larger than some characteristic length $\xi_{1}$.
%, which may be thought as the entanglement size analog in the liner chain solution.
Individual rings in concentrated solutions of small molecular weight $N<g_{1} \equiv (\xi_{1}/a)^2 \phi^{1/4}$ ($N$ is the number of monomers in each ring, $a$ is the monomer size and $\phi$ is the monomeric volume fraction) thus show Gaussian behaviors with the size $R_0 \simeq \xi (N/g)^{1/2} \simeq a N^{1/2} \phi^{-1/8}$ where $\xi \simeq a g^{3/5} \simeq a \phi^{-3/4}$ is the correlation length of concentration fluctuation.
In the present paper, we introduce the notion of the {\it topological volume fraction} and construct a mean-filed theory for the concentrated solution of noncatenated long ring polymers. Requiring the theory to be compatible with the above feature associated with the length scale $\xi_{1}$, we show that the scaling exponent for the long chain limit is given by $\nu=1/3$. However, extremely long chains $N>N^{*}= C g_{1}$ with a large numerical factor $C \simeq 10 $ are required for this asymptotic to be reached, thus, most practical cases fall into a broad crossover region ($g_{1} < N < N^*$) where the apparent exponent is given by $\nu=2/5$.
At various stage of the paper, we will look into the physics behind the C-D theory in the light of the present argument.
%By comparing the C-D theory to the present one at various stages of the paper, we will look into the physics behind the C-D conjecture, and point out its drawback.
We also suggest a connection to the CG model through a {\it topological blob} which adjusts its own size in an intriguing way in the crossover region.
This view provides a natural bridge between otherwise conflicting two lines of previous conjectures (C-D and CG).

{\it Mean field theory}---
A rigorous analysis  of the topological effects may rely on the mathematical knot invariance, which, however, seems to be yet formidable.
We instead seek for a physically motivated coarse-grained phenomenological description.
A successful prototype of such an approach is already found in the dynamics and rheology of high molecular weight linear polymer melts in which the tube model provides a bridge between molecular level topological constraints and macroscopic material properties~\cite{deGennes, Doi_Edwards}.
In the present problem, we also postulate an intrinsic length scale $\xi_{1}$ which is an analogue of the tube diameter in the linear chain solutions, below which the topological effects are irrelevant.
At this stage, what is needed is an element to treat large scale {\it static} behaviors of unconcatenated rings, just as the reptation theory provides a basic framework to treat the dynamical properties of entangled linear chain solutions beyond the tube size.

A key observation from previous studies is that the topological constraints could be effectively represented as excluded-volume effects. 
Notable topics in this line include an entropic repulsion between untangled rings~\cite{Frank-Kamenetskii}, a topological swelling of isolated random knots~\cite{topological_swelling1, topological_swelling2} and the anomalous bond correlation function of planar rings~\cite{Sakaue_Wada}. 
It is expected that the above conjecture, first proposed by des Cloizeaux~\cite{desCloizeaux}, would be equally useful for the present problem, too.

Let us consider a concentrated solution of ring polymers. 
The equilibrium spatial size $R$ (such as a radius of gyration) of each ring is a function of $N$ and the monomer volume fraction $\phi$.
The number of total monomers in the region of volume $\sim R^3$ of single ring is $\sim R^3 \phi/a^3$, thus, the number of rings there is
\begin{eqnarray}
N_R \simeq \frac{R^3 \phi}{a^3 N} 
\label{N_x}
\end{eqnarray}
Using a ``self density" $\phi_s = a^3 N/R^3$, it can be written as $N_R = \phi/\phi_s$.

%In addition to the usual osmotic pressure present in linear chain solutions (given in eq.~(\ref{Osm_linear})), there arises an excess contribution due to the topological effect.
The topological contribution to the free energy consists of terms arising from (i) noncatenation constrains among different rings and (ii) intra-ring effects associated with self-knotting~\cite{C-D}:
\begin{eqnarray}
F = F_{inter} + F_{intra}
\label{free_energy_basis}
\end{eqnarray}
In C-D theory, the first term is evaluated by assigning $\sim k_BT$ to each of $N_R$ neighboring rings, leading to the estimate $F_{inter}/k_BT \simeq (R^3 \phi)/(a^3 N)$.
One may notice that this amounts to the evaluation of binary interactions.
In concentrated solutions of long rings, however, the many-body correlation effects should become progressively important, which may be taken into account in line with the excluded-volume analogy as followings.
The effective excluded-volume $v_R$ of the ring with the spatial size $R$ scales its volume
\begin{eqnarray}
v_R\simeq R^3 Y
\end{eqnarray}
where a dimensionless factor $Y$ (independent of $N$) shall be fixed later to be consistent with the presence of the length scale $\xi_{1}$ mentioned earlier.
To evaluate the resultant repulsive interaction, we adopt a classical van der Waals theory of fluids whose free energy density for a one component fluid with the volume fraction $\phi$ and the excluded volume $v$ is given by $f(\phi)/(k_BT) = [\phi/v] \ln{[\phi/(1-\phi)]} -\epsilon \phi^2$.
%\begin{eqnarray}
%\frac{f(\phi)}{k_B T } = \frac{\phi}{v} \ln{\frac{\phi}{1-\phi}}-\epsilon \phi^2
%\end{eqnarray}
In our case, this may be transformed as $f_{inter}/(k_BT) \simeq -[\phi_R/v_R] \ln{[(1-\phi_R)]}$
%\begin{eqnarray}
%\frac{f_{inter}}{k_B T } \simeq - \frac{\phi_R}{v_R}  \ln{(1-\phi_R)}
%\end{eqnarray}
where we set $\epsilon=0$ (athermal), and as usual for the polymer solution theory, the ideal gas term is irrelevant here~\cite{Grosberg_Khokhlov}, and
\begin{eqnarray}
\phi_R \equiv \frac{v_R N_R}{R^3}= N_R Y
\label{phi_R}
\end{eqnarray}
is the ring's ``volume fraction" of topological origin.
The free energy per ring $F_{inter} =f_{inter} \times R^3/N_R$  is thus
\begin{eqnarray}
\frac{F_{inter}(\phi_R)}{k_B T } = - \ln{(1-\phi_R)}
\label{F_topo_G}
\end{eqnarray}
 An inspection of eq.~(\ref{F_topo_G}) indicates that the larger size $R$ costs free energy due to the noncatenation constraint. 
Thus, this topology effect leads to the squeezing of the ring toward a globular state, which should be negotiated with the unknotting constraint. A scaling analysis suggests that this free energy cost for the confinement should be written as
\begin{eqnarray}
\frac{F_{intra}}{k_BT}  \simeq \left( \frac{R_0}{R}\right)^{\delta} =  \left( \frac{a N^{1/2} \phi^{-1/8}}{R}\right)^{\delta} \simeq \left( \frac{N Y^2 \phi^{5/4}}{\phi_R^2}\right)^{\delta/6}
\label{F_conf_G}
\end{eqnarray}
where $R_0 = \xi (N/g)^{1/2}=a N^{1/2} \phi^{-1/8}$ is the size of unperturbed Gaussian ring (more precisely, the Gaussian ring of the blobs of size $\xi = a g^{3/5}= a\phi^{-3/4}$), and use has been made of eqs.~(\ref{N_x}) and~(\ref{phi_R}) in the last equality. 
One then requires $F_{intra}$ to be extensive ($\sim N$) with $\phi_R$ given, which leads to $\delta=6$~\footnote{It is equivalent to say that avoiding knots would be a bulk constraint, i.e., we require $F_{intra} \rightarrow k F_{intra}$ upon enlarging the system size as $N \rightarrow kN$ and $V \simeq R^3 \rightarrow k V$, which leads to $\delta=6$.};
\begin{eqnarray}
\frac{F_{intra}(\phi_R; \ N)}{k_BT} = \phi_R^{-2} NY^2 \phi^{5/4}
\label{F_conf_G_2}
\end{eqnarray}
Note that this differs from a naive guess $\delta=2$ adopted in the C-D theory which, however, is no longer correct for non-ideal chains confined in closed cavity (see ref.~\cite{Sakaue_Raphael} for a general discussion on it).
Indeed, eq.~(\ref{F_conf_G_2}) is naturally expected in the context of the excluded-volume analogy~\cite{Grosberg_Khokhlov, Sakaue_Raphael} and applies when rings feel squeezing $R<R_0 \Leftrightarrow N>g_1$.
%Using eqs.~(\ref{N_x}) and~(\ref{phi_R}), eq.~(\ref{F_conf_G}) can be rewritten as
It is noted that while $F_{inter}$ solely depends on the topological volume fraction $\phi_R$, $F_{intra}$ linearly increases with the chain length $N$ at a given $\phi_R$. This reflects different aspects of the constraints between nonconcatenation and unknotting.

\paragraph{Short scale behavior}

\begin{figure}[h]
\includegraphics[width=0.45\textwidth]{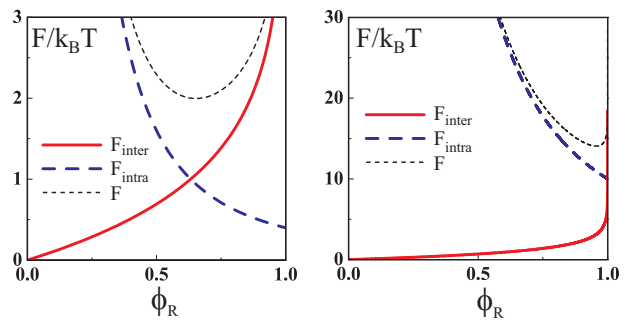}
\caption{Plots of free energy as a function of $\phi_R$ for short ring ${\mathcal B} N/g_{1}=0.4$ (left) and long ring ${\mathcal B} N/g_1=10$ (right). }
\label{Fig1}
\end{figure}

Figure~\ref{Fig1} shows typical free energy profiles for short and long rings. 
For small $N$, the free energy minimum is attained at low $\phi_R < 1$.
One may find a situation $F_{inter}({\tilde \phi_R}) \simeq F_{intra}({\tilde \phi_R}; \ g_{1}) \simeq k_BT$ at ${\tilde \phi_R}$.
This indicates that, in short length scale $r< \xi_{1}= a g_{1}^{1/2}\phi^{-1/8}$, the topological effect is not effective, and the ideal ring behavior is observed.
From the above condition, we find
\begin{eqnarray}
Y \simeq g_{1}^{-1/2} \phi^{-5/8} \  {\mathcal B}^{1/2}({\tilde \phi}_R)
\label{eq:Y}
\end{eqnarray}
where ${\mathcal B}({\tilde \phi}_R) \simeq  {\tilde \phi}_R^2  \simeq 0.5 $.
We shall shortly argue that this numerical factor ${\mathcal B}$ is crucial for the characterization of the crossover regime.
The confinement free energy eq.~(\ref{F_conf_G_2}) can now be rewritten as
\begin{eqnarray}
\frac{F_{intra}(\phi_R; N)}{k_BT} = {\mathcal B} \phi_R^{-2} \frac{N}{g_{1}} = \phi_R^{-2} \frac{N}{g_2}
\label{F_conf_G_3}
\end{eqnarray} 
where $g_2 \equiv g_{1}/{\mathcal B}$.

\paragraph{Free energy minimization}
Now let us seek for the spatial structure of rings in large scales $N > g_{1}$.
Here, the energy scale is much larger than the thermal energy (see Fig.~\ref{Fig1} ), thus, the equilibrium size would be readily determined through the free energy minimization with respect to $\phi_R$. 
%Minimization of the free energy with respect to $\phi_R$ reads
%\begin{eqnarray}
%(1-\phi_R)^{-1} = 2 {\mathcal B} \phi_R^{-3} \frac{N}{g_{1}}
%\end{eqnarray}
%%%%%%%%%%%%%%%%%%%%%%%%%%%%%%%%%%%%%%%%%%%%%%%%%%%%%
\begin{figure}[h]
\includegraphics[width=0.38\textwidth]{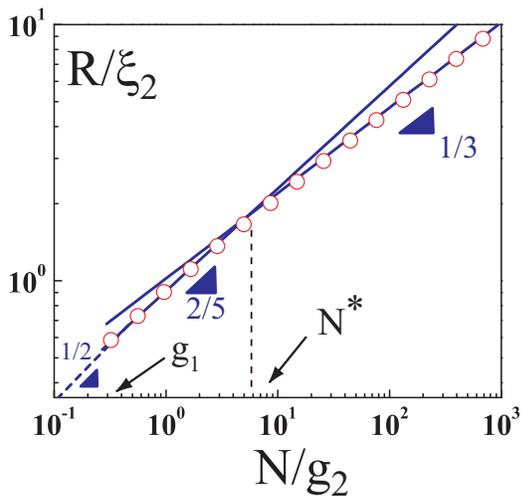}
\caption{Normalized plot of the ring size $R/\xi_2$ as a function of the chain length $N/g_2$ (double logarithmic scale).  }
\label{Fig2}
\end{figure}
%%%%%%%%%%%%%%%%%%%%%%%%%%%%%%%%%%%%%%%%%%%%%%%%%%%%%
The result of numerical solution shown in Fig.~\ref{Fig2} is summarized as follows; (i) sufficiently long rings ($N>N^*$) obey a compact statistics with $\nu=1/3$, (ii) rings with intermediate length ($g_{1}<N<N^*$) are well characterized by the effective exponent $2/5$.
Therefore, one may say that the statistics of concentrated ring polymers has well-defined two regimes (ideal and compact statistics for short $N<g_{1}$ and long $N>N^*$ rings, respectively) which are separated by a wide crossover region spanning over about one order of magnitude.
%%%%%%%%%%%%%%%%%%%%%%%%%%%%%%%%%%%%%%%%%%%%%%%%%%%%%
\begin{figure}[h]
\includegraphics[width=0.34\textwidth]{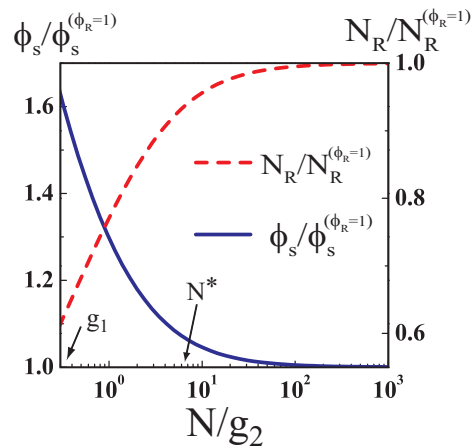}
\caption{Dependence of the self density $\phi_s$ and the number of neighboring rings $N_R$ on the chain length $N$ (semi-logarithmic plot). $\phi_s$ decreases from $\phi_s \simeq \phi^{3/8} g_1^{-1/2}$ at $N=g_1$ toward $\phi_s^{(\phi_R=1)} \simeq \phi^{3/8} g_2^{-1/2}$ at $N>N^*$, whereas $N_R$ increases from $N_R \simeq \phi^{5/8} g_1^{1/2}$ at $N=g_1$ toward $N_R^{(\phi_R=1)} \simeq \phi^{5/8} g_2^{1/2}$ at $N>N^*$.}
\label{Fig3}
\end{figure}
%%%%%%%%%%%%%%%%%%%%%%%%%%%%%%%%%%%%%%%%%%%%%%%%%%%%%
To get an insight into this crossover, we note that from eqs.~(\ref{N_x}),~(\ref{phi_R}),~(\ref{F_topo_G}),~(\ref{eq:Y}),(\ref{F_conf_G_3}), a mean-field solution for $N>g_{1}$ can be generally written in the form
\begin{eqnarray}
R= \xi_{2} \left(\frac{N \phi_R^{(eq)}(N/g_2)}{g_{2}}\right)^{1/3} =\xi_{\natural}(N) \left(\frac{N}{g_{\natural}(N)}\right)^{1/3} 
\end{eqnarray}
where $\xi_2 \equiv a \phi^{-1/8} g_{2}^{1/2}  = \xi_1 {\mathcal B}^{-1/2}$and $\phi_R^{(eq)}$ denotes the equilibrium value which depends solely on $N/g_2$.
This indicates that upon renormalization a large scale spatial organization of individual rings is formally characterized by a space-filling curve with fractal dimension $\nu^{-1} = 3$.
Unlike usual cases, however, a renormalization factor, which is invoked in an earlier study as {\it topological blob}~\cite{crumpled_globule}, $\xi_{\natural}(N) = a g_{\natural}^{1/2}(N) \phi^{-1/8}$ is not constant, but rather depends on $N$, i.e., starting from $g_{\natural}=g_1$ at $N=g_1$, it slowly (logarithmically) increases toward $g_{\natural}=g_2$ at $N=N^*$, then saturates.
It is this slow evolution toward a topological dense-packed limit ($\phi_R \rightarrow 1$) that renders a small excess correction to the exponent in the large crossover region.
Such a feature is demonstrated in Fig.~\ref{Fig3} where the self-density $\phi_s$ and $N_R$ (roughly corresponding to the number of neighboring rings) are plotted against the chain length.
A point $N=N^*$ at which these observables saturate corresponds to the onset of the compact statistics, i.e.,
 $\phi_R \rightarrow 1$ at $N > g_{2}$ which leads to the onset length $N^{*} = C g_{1}$ with $C \simeq 10 $ (Fig.~\ref{Fig3}).
To get a number, let us adopt an entanglement length as a measure of $g_1$, and substitute a typical value $\sim 100$. This yields $N^{*} \sim 1000$~\footnote{We also have an estimate from eqs.~(\ref{phi_R}),~(\ref{eq:Y}) that $N_R^{(\phi_R=1)} = Y^{-1} \simeq g_2^{1/2}\phi^{5/8} \simeq 15$ for melts ($\phi \simeq 1$) in good agreement with a recent simulation~\cite{Vettorel}.}.

Several remarks are now in order.
As a mean-field theory, if appropriately constructed, is expected to provide an accurate description on concentrated solutions, it would be of value to get another perspective on the structure of free energies.
First, eq.~(\ref{F_topo_G}) for $F_{inter}$ can be expanded in a virial series $F_{inter}/k_BT \simeq \phi_R + \phi_R^2/2 + \cdots $. Retaining only a lowest order term (second virial approximation), one finds $F_{inter}/k_BT \simeq R^3 \phi Y/(a^3 N)$ which corresponds to the free energy adopted in the C-D theory aside from a factor $Y$~\footnote{Indeed, it is nontrivial that topological interactions among rings could be described by excluded-volume proportional to the volume of the ring even in melts, and this series expansion would be helpful in comparison with C-D theory. See a footnote (4) in ref.~\cite{C-D}.}.
Our analysis suggests, however, that higher order terms are essential. 
Second, a view of the compact ring of size $R \ (<R_0)$ as a dense piling of topological blob implies $\xi_{\natural} \simeq \xi (g_{\natural}/g)^{1/2}$ and $a^3 g_{\natural}/\xi_{\natural}^3 = \phi_s$. One thus finds $\xi_{\natural} \simeq a \phi^{1/4} \phi_s^{-1} \simeq a \phi^{-3/4} N_R$ and $g_{\natural} \simeq \phi^{-5/4} N_R^2$.
% = a \phi^{-3/4} \phi_R/Y = a \phi^{-1/8}g_1^{1/2} \phi_R {\mathcal B}^{-1/2}$. 
Assigning $\sim k_BT$ per topological strand, we obtain $F_{intra}/(k_BT) \simeq N/g_{\natural} \simeq  R^3/\xi_{\natural}^3$, which coincides with the scaling derivation of $F_{intra}$ (eq.~(\ref{F_conf_G_2}))~\cite{Sakaue_Raphael}. 
Moreover, the deduced relations $\xi_{\natural} \simeq \xi N_R$ and $g_{\natural} \simeq g N_R^2$ indicate that the number of units (correlation blobs) required to form a topological constraint is on the order of $\sim N_R^2$.

%Under that condition, its approximate solution is
%\begin{eqnarray}
%\phi_R &\simeq& 1-  \left( \frac{g_{1}}{{\mathcal B} N}\right) \nonumber \\
%\Leftrightarrow
%R &\simeq& a N^{1/3} g_{1}^{1/6}\phi^{-1/8} {\mathcal B}^{-1/6}\left[ 1-  \left( \frac{ g_{1}}{{\mathcal B}N}\right)\right]^{1/3} \nonumber \\
%&\simeq&\xi_{1}\left(\frac{N}{g_{1}}\right)^{1/3} {\mathcal B}^{-1/6}\left[ 1-  \left( \frac{ g_{1}}{{\mathcal B}N}\right)\right]^{1/3}
%\label{X_scaling}
%\end{eqnarray}
The optimum free energy for large $N (\gg g_1)$  is
$F/(k_B T) \simeq N/g_{\natural}$
, i.e., an order of the magnitude of thermal energy per topological strand.
This translates into the osmotic pressure
$\Pi_{top} \simeq F N_R/R^3 \simeq k_B T \phi/(a^3 g_{\natural}) \simeq (N_R)^{-2} k_BT \phi^{9/4}/a^3$.
%Most theoretical models of polymer solution in good solvents predict the entanglement spacing to be proportional to the correlation length $\xi= a \phi^{-3/4}$. Following this, we also assume $\xi_{2} \sim \xi$, or, in terms of the number of segments in the topological strand, $g_{2} = {\mathcal C}_0 g \simeq {\mathcal C}_0 \phi^{-5/4}$, where we have explicitly include a numerical factor ${\mathcal C}_0 \gg 1$ to indicate that a number of units (correlation blobs) are required to form an entanglement (topological constraint in the present contest). 
%This leads to
%\begin{eqnarray}
%\Pi_{top} \simeq {\mathcal C}_0^{-1} \frac{k_B T \phi^{9/4}}{a^3}
%\end{eqnarray}
Adding this to the usual excluded-volume contribution $\Pi_{linear} \simeq k_BT \phi^{9/4}/a^3$~\cite{deGennes}, the total osmotic pressure thus can be written as
\begin{eqnarray}
\Pi = (1+N_R^{-2}) \Pi_{linear}
\end{eqnarray}
where one may identify the small factor as the increase of the local excluded-volume effect, i.e., $N_R^{-2} \leftrightarrow \delta v/v$ in accordance to the des Cloizeaux conjecture~\cite{desCloizeaux}.
%\footnote{Let us denote the excluded volume as $v$ so that $\phi = c v$. Then, the osmotic pressure is $\Pi \simeq \frac{k_BT}{\xi^3} \simeq v^{-1} \phi^{9/4} = v^{5/4}c^{9/4}$. Let us slightly increase the excluded volume as $v \rightarrow v+ \delta v$ ($\delta v/v \ll 1$). Then the increase in the osmotic pressure is $\delta \Pi = \Pi(v+\delta v)-\Pi(v) \simeq \Pi(v) \frac{\delta v}{v}$.}.

To summarize, the present attempt strongly suggests the applicability of the excluded-volume analogy to the problem of dense ring solutions.
It sets up a way to handle immense topological constraints, thus allowing one to capture the essential static properties in such systems.
In particular, it provides us with a fairly accurate description of the ring dimension over the range of semidilute to concentrated solution regimes.
Other key quantities such as $\phi_{s}$, $N_R$ as well as the estimated onset length $N^*$ of the compact statics are also in line with reported numerical observations~\cite{Muller1, Brown, Muller2, Muller3,Vettorel,Suzuki}.
%To summarize, we have presented a mean-field theory for concentrated solutions of ring polymers inspired by the excluded-volume analogy. 
It should be noticed that our starting point is similar in spirit to that in C-D theory, i.e., the identification of two competing contributions in eq.~(\ref{free_energy_basis}). Rather, it is the difference in the basic structure of the free energy that leads to deep insights into the hierarchical spatial structure of ring solutions, hence a bridge to the CG concept.
Further careful studies are awaited to examine the validity and the limitation of the present phenomenological approach as well as the usefulness of the concept such as the topological volume fraction, etc.
As stated in ref.~\cite{Vettorel}, there are various situations in molecular systems, i.e., the collapse of a gel, the existence of chromosome territories, the compatibility enhancement~\cite{Ring_linear_compatibility} etc. where topological constraints matter.
We hope that the present analysis provides valuable insights into such cases, too.
T.S. thanks H. Nakanishi and A. Yoshimori for useful comments and discussions.

%\acknowledgments 
%T.S. thanks H. Nakanishi and A. Yoshimori for useful comments and discussions. 

\end{document}